\documentclass[12pt]{article}
\usepackage{a4}
\usepackage{amsfonts}
\usepackage{amsmath,amssymb}
\usepackage{graphicx}
\usepackage{slashed}
\usepackage{amsmath}

\usepackage[english]{babel}

\usepackage{multirow}
\usepackage[margin=0.8in]{geometry}

\newcommand{\overbar}[1]{\mkern1.5mu\overline{\mkern-1.5mu#1\mkern-1.5mu}\mkern 1.5mu}

\makeatother

\def\un#1{\relax\ifmmode\@@underline#1\else
        $\@@underline{\hbox{#1}}$\relax\fi}

\let\du=\du                     

\def\b{\beta}

\def\f{\phi}
\def\g{\gamma}

\def\j{\psi}

\def\l{\lambda}
\def\m{\mu}
\def\n{\nu}

\def\p{\pi}

\def\G{\Gamma}

\def\bo{{\raise-.3ex\hbox{\large$\Box$}}}               
\def\TH{{\raise.2ex\hbox{$\displaystyle \bigodot$}\mskip-4.7mu \llap H \;}}
\def\face{{\raise.2ex\hbox{$\displaystyle \bigodot$}\mskip-2.2mu \llap {$\ddot
        \smile$}}}                                      

\def\VEV#1{\left\langle #1\right\rangle}        
\def\abs#1{\left| #1\right|}                    
\def\leftrightarrowfill{$\mathsurround=0pt \mathord\leftarrow \mkern-6mu
        \cleaders\hbox{$\mkern-2mu \mathord- \mkern-2mu$}\hfill
        \mkern-6mu \mathord\rightarrow$}
\def\dvec#1{\vbox{\ialign{##\crcr
        \leftrightarrowfill\crcr\noalign{\kern-1pt\nointerlineskip}
        $\hfil\displaystyle{#1}\hfil$\crcr}}}           
\def\dt#1{{\buildrel {\hbox{\LARGE .}} \over {#1}}}     

\def\frac#1#2{{\textstyle{#1\over\vphantom2\smash{\raise.20ex
        \hbox{$\scriptstyle{#2}$}}}}}                   
\def\sfrac#1#2{{\vphantom1\smash{\lower.5ex\hbox{\small$#1$}}\over
        \vphantom1\smash{\raise.4ex\hbox{\small$#2$}}}} 
\def\bfrac#1#2{{\vphantom1\smash{\lower.5ex\hbox{$#1$}}\over
        \vphantom1\smash{\raise.3ex\hbox{$#2$}}}}       
\def\afrac#1#2{{\vphantom1\smash{\lower.5ex\hbox{$#1$}}\over#2}}    

\def\[{\lfloor{\hskip 0.35pt}\!\!\!\lceil}
\def\]{\rfloor{\hskip 0.35pt}\!\!\!\rceil}

\def\du#1#2{_{#1}{}^{#2}}

\def\un{\underline}
\def\fracmm#1#2{{{#1}\over{#2}}}

\def\low#1{{\raise -3pt\hbox{${\hskip 0.75pt}\!_{#1}$}}}

\def\Dot#1{\buildrel{_{_{\hskip 0.01in}\bullet}}\over{#1}}
\def\dt#1{\Dot{#1}}

\newskip\humongous \humongous=0pt plus 1000pt minus 1000pt

\newif\ifdtup

\def\({\left(}
\def\){\right)}

\def\beq{\begin{equation}}
\def\eeq{\end{equation}}
\def\bea{\begin{eqnarray}}
\def\eea{\end{eqnarray}}

\newcommand{\be}{\begin{equation}}
\newcommand{\ee}{\end{equation}}
\newcommand{\nbe}{\begin{equation*}}
\newcommand{\nee}{\end{equation*}}

\newcommand{\lb}{\label}

\begin{document}

\thispagestyle{empty}

{\hbox to\hsize{
\vbox{\noindent April 2018 \hfill IPMU17-0101 \\
revised version}}}
{\hbox to\hsize{
\vbox{\noindent  \hfill }}}
\noindent
\vskip2.0cm
\begin{center}

{\Large\bf Gravitino and Polonyi production in supergravity
}

\vglue.3in

Andrea Addazi~${}^{a}$, Sergei V. Ketov~${}^{b,c,d,e}$ and Maxim Yu. Khlopov~${}^{f,g}$
\vglue.1in

${}^{a}$~Department of Physics, Fudan University, 220 Handan Road, 200433 Shanghai, China\\
${}^b$~Department of Physics, Tokyo Metropolitan University, \\
Minami-ohsawa 1-1, Hachioji-shi, Tokyo 192-0397, Japan\\
${}^c$~Institute of Physics and Technology, Tomsk Polytechnic University,\\
30 Lenin Ave., Tomsk 634050, Russia \\
${}^d$~International Centre for Theoretical Physics and South American Institute
for Fundamental Research,  Rua Dr. Bento Teobaldo Ferraz, 271 – Barra Funda, Sao Paulo, CEP 01140-070 Brazil\\
${}^e$~Kavli Institute for the Physics and Mathematics of the Universe (IPMU),
\\The University of Tokyo, Chiba 277-8568, Japan \\
${}^f$~Centre for Cosmoparticle Physics Cosmion; National Research Nuclear University MEPHI (Moscow Engineering Physics Institute), Kashirskoe Sh., 31, Moscow 115409, Russia \\
${}^g$~APC Laboratory 10, rue Alice Domon et L\'eonie Duquet, 75205 Paris Cedex 13, France

\vglue.1in
3209728351@qq.com, ketov@tmu.ac.jp, khlopov@apc.in2p3.fr
\end{center}

\vglue.3in

\begin{center}
{\Large\bf Abstract}
\end{center}
\vglue.1in
\noindent  We study production of gravitino and Polonyi particles in the minimal 
Starobinsky-Polonyi $\mathcal{N}=1$ supergravity with inflaton belonging to a massive vector supermultiplet.  Our model has only one free parameter given by the scale of spontaneous SUSY breaking triggered by Polonyi chiral superfield. The vector supermultiplet generically enters the action non-minimally, via an arbitrary real function. This function is chosen to generate the inflaton scalar potential of the Starobinsky model. Our supergravity model can be reformulated as an abelian supersymmetric  gauge theory with the vector gauge superfield coupled to two (Higgs and Polonyi) chiral superfields interacting with supergravity, where the $U(1)$ gauge symmetry is spontaneously broken. We find that Polonyi and gravitino particles are efficiently produced during inflation, and estimate their masses and the reheating temperature. After inflation, perturbative decay of inflaton also produces Polonyi particles that rapidly decay into gravitinos. As a result, a coherent picture of inflation and dark matter emerges, where the abundance of produced gravitinos after inflation  fits the CMB constraints as a Super Heavy Dark Matter (SHDM) candidate. Our scenario avoids the notorous gravitino and Polonyi problems with the Big Bang Nucleosynthesis (BBN)  and DM overproduction.

\newpage

\section{Introduction}

The Planck data \cite{Ade:2015xua, Ade:2015lrj,Array:2015xqh} of the Cosmic Microwave Background (CMB) radiation favors slow-roll single-large-field chaotic inflation with an approximately flat plateau of the scalar potential, driven by single inflaton (scalar) field. The simplest geometrical realization of this description is provided by the Starobinsky model
\cite{Starobinsky:1980te}. It strongly motivates us to connect this class of inflationary models to particle physics theory beyond the Standard Model (SM) of elementary particles. A reasonable way of theoretical realization of this program is via embedding of the inflationary models  into $\mathcal{N}=1$ supergravity.  It is also the first natural step towards 
unification of inflation with the Supersymmetric Grand Unified Theories (SGUTs)  and string theory. Inflaton is expected to be mixed with other scalars, but this mixing has to be small.
The inflationary model building in the supergravity  literature is usually based on an assumption that inflaton belongs to a chiral (scalar) supermultiplet, see  e.g.,  the reviews \cite{Yamaguchi:2011kg,Ketov:2012yz} for details.  However,  inflaton can also belong to  a massive $\mathcal{N}=1$ {\it vector} multiplet instead of a chiral one.  Since there is only one real scalar in a massive $\mathcal{N}=1$ vector multiplet, there is no need of stabilization of its (scalar) superpartners, and   the $\eta$-problem does not exist because   the scalar potential of a vector multiplet is given by the $D$-term instead of the $F$-term.   The minimal supergravity models with inflaton belonging to a massive vector multiplet were proposed in  Refs.~\cite{Farakos:2013cqa,Ferrara:2013rsa} by using the non-minimal self-coupling of a vector multiplet, paramaterized by an arbitrary real function \cite{VanProeyen:1979ks}.  These models can accommodate any desired values of the CMB observables (the scalar tilt $n_s$ and the tensor-to-scalar ratio $r$), because the corresponding single-field (inflaton) scalar potential is given by the derivative squared of that (arbitrary) real function. However, all models of Refs.~\cite{Farakos:2013cqa,Ferrara:2013rsa} have the vanishing vacuum energy after inflation, i.e. the vanishing cosmological constant, and the vanishing vacuum expectation value (VEV) of the auxiliary fields, so that supersymmetry (SUSY) is restored after inflation and only a Minkowski vacuum is allowed.  A simple extension of the models \cite{Farakos:2013cqa,Ferrara:2013rsa} was proposed in 
Refs.~\cite{Aldabergenov:2016dcu,Aldabergenov:2017bjt}, where 
a {\it Polonyi} (chiral) superfield with a linear superpotential \cite{Polonyi:1977pj} was added to the action,  leading to a spontaneous SUSY breaking and a {\it de-Sitter} vacuum after inflation.

A successful theoretical embedding of inflation into supergravity models  is, clearly, a necessary but is not a sufficient condition. Even when these models are well compatible with the Planck constraints on the ($r,n_{s}$), they
still may (and often, do) lead to  incompatibility with the (Hot and Cold) Dark Matter (DM) abundance and the Big Bang Nucleosynthesis (BBN). A typical issue is known as the {\it gravitino problem}:  in order to not 
ruin the BBN,  gravitinos  must not decay in the early thermal bath injecting  hadrons or radiation during the BBN epoch \cite{Khlopov:1984pf,Khlopov:1993ye,Kawasaki:2004qu,Khlopov:2015oda}.  As is well  known, the BBN is very sensitive to initial conditions,  while each extra hadron or radiation can radically jeopardize the BBN picture, leading to disastrous 
incompatibility with cosmological and astrophysical data. In addition, the so-called {\it Polonyi problem} was also pointed out in the literature:  Polonyi particles decay can also jeopardize the success of the BBN 
 \cite{Banks:1993en,deCarlos:1993wie,Coughlan:1983ci,Moroi:1994rs,Kawasaki:1995cy,Moroi:1999zb}. Indeed,
 a generic supergravity model  predicts a disastrous overproduction of gravitinos and/or Polonyi particles or neutralinos. Any specific predictions are model-dependent, because they are very sensitive to  the mass spectrum and the  
 parameter range under consideration. The mass pattern selects the  leading production mechanism or channel: either thermal (WIMP-like) production or/and non-thermal production sourced by inflation and decays of other heavier 
 particles. The last channel includes a possible (different) production mechanism due to evaporating Primordial Black Holes (PBH's)  that may be formed in the early Universe \cite{Khlopov:1985jw,Khlopov:2004tn,Khlopov:2008qy}. Our minimal estimation of the probability of the mini-PBH's formation at the long dust-like preheating stage after inflation gives such a low value that the successive evaporation of the mini-PBH's doesn't lead to a significant contribution to gravitino production (Sec.~3). However, if inflation ends by a first order phase transition, the situation  drastically changes, and copious production of mini-PBHs in bubble collisions \cite{miniPBH,miniPBH2} can lead to a huge gravitino overproduction, thus excluding the first order phase transition exit from inflation.
 
We consider the very specific, minimalistic and, hence, a bit oversimplified   $\mathcal{N}=1$ supergravity model of inflation, with inflaton belonging to a massive vector multiplet. We demonstrate that our model avoids the overproduction and BBN problems,  while it naturally accounts for the right amount of cold DM. We assume both Polonyi field,  triggering a spontaneous SUSY breaking at high scales, and the massive gravitino, produced during inflation, to be super-heavy, and call it the Super-Heavy Gravitino Dark Matter (SHGDM) scenario.  A production of super-heavy scalars during inflation was first studied in Refs.~\cite{Chung:1998zb,Chung:2001cb}, whereas the gravitino production sourced by inflation  was considered in Refs.~\cite{Addazi:2016mtn,Addazi:2016bus,Addazi:2017kbx}, though without specifying a particular model.  In this paper we apply the methods of  Refs.~\cite{Addazi:2016mtn,Addazi:2016bus,Addazi:2017kbx} to the specific
Starobinsky-Polonyi supergravity model proposed in Refs.~\cite{Aldabergenov:2016dcu,Aldabergenov:2017bjt}. The supersymmetric partners of known particles (beyond the ones present in the model) are assumed to be heavier than Polonyi and gravitino particles (in the context of High-scale SUSY), also in order to overcome several technicalities in our calculations. Some of the physical predictions of our model are (i) the Polonyi mass is a bit higher than two times of  the gravitino mass, and (ii) the inflaton mass is slightly higher than two Polonyi masses. This implies that inflaton can decay into Polonyi particles that, in their turn, decay into a couple of gravitinos.  We show that super-heavy gravitinos produced from inflation  and Polonyi decays can fit the cold DM abundance. The parameter spaces of inflation and  cold DM  are thus  linked to each other  in a coherent unifying picture. 
 
Our paper is organized as follows. In Sec.~2 we review our model.   In Sec.~3 we consider the gravitino and Polonyi particle production mechanisms. Sec.~4 is our conclusion and outlook.
\vglue.2in

\section{The model}

In this Section we briefly review  the inflationary model of Refs.~\cite{Aldabergenov:2016dcu,Aldabergenov:2017bjt}. We use  the natural units with the reduced Planck mass $M_{Pl}=1$.~\footnote{Our notation and conventions coincide with the standard ones in Ref.~\cite{Wess:1992cp}, including the spacetime signature $(-,+,+,+)$. The $N=1$ superconformal calculus \cite{Ferrara:2013rsa,VanProeyen:1979ks} after the superconformal gauge fixing is equivalent to the curved superspace description of $N=1$ supergravity that is used here.}

The model has two chiral superfields $(\Phi, H)$ and a real vector superfield $V$, all coupled to supergravity, and having the Lagrangian 
\begin{equation}
\label{hsslag}
\mathcal{L}=\int d^2\theta 2\mathcal{E}\left\lbrace \frac{3}{8}(\overbar{\mathcal{D}}\overbar{\mathcal{D}}-8\mathcal{R})e^{-\frac{1}{3}(K+2J)}+\frac{1}{4}W^\alpha W_\alpha +\mathcal{W}(\Phi) 
\right\rbrace +{\rm h.c.}~,
\end{equation}
in terms of the chiral scalar curvature superfield $\mathcal{R}$, the chiral density superfield $\mathcal{E}$ and the superspace covariant spinor derivatives  $(\mathcal{D}_{\alpha}, \overbar{\mathcal{D}}^{\dt{\alpha}})$,  a K\"ahler potential 
$K= K(\Phi,\overbar{\Phi})$ and a superpotential $\mathcal{W}(\Phi)$, the abelian (chiral) superfield strength
$W_\alpha\equiv-\frac{1}{4}(\overbar{\mathcal{D}}\overbar{\mathcal{D}}-8\mathcal{R})\mathcal{D}_\alpha V$,
and a real function $J=J(He^{2gV}\overbar{H})$ with the coupling constant $g$.
   
The Lagrangian \eqref{hsslag} is invariant under the supersymmetric $U(1)$ gauge transformations
\begin{gather}
H\rightarrow H'=e^{-igZ}H~,\;\;\;\overbar{H}\rightarrow \overbar{H}'=e^{ig\overbar{Z}}\overbar{H}~,\label{supergh}\\
V\rightarrow V'=V+\frac{i}{2}(Z-\overbar{Z})~,\label{supergv}
\end{gather}
whose gauge parameter $Z$ itself is a chiral superfield. 

The chiral superfield $H$ can be gauged away via the gauge fixing of these transformations by a gauge condition $H=1$. Then the Lagrangian \eqref{hsslag} gets simplified to
\begin{equation} \label{sslag}
\mathcal{L}=\int d^2\theta 2\mathcal{E}\left\lbrace \frac{3}{8}(\overbar{\mathcal{D}}\overbar{\mathcal{D}}-8\mathcal{R})e^{-\frac{1}{3}(K+2J)}
+\frac{1}{4}W^\alpha W_\alpha +\mathcal{W} \right\rbrace +{\rm h.c.}
\end{equation}

After eliminating the auxiliary fields and moving from the initial (Jordan) frame to the Einstein frame, the {\it bosonic} part of the Lagrangian  (\ref{sslag}) reads \cite{Aldabergenov:2016dcu} ~\footnote{The primes and capital latin subscripts denote the derivatives with respect to the corresponding fields.}
 \begin{equation} \label{complag}
e^{-1}\mathcal{L}=-\frac{1}{2}R-K_{A\bar{A}}\partial_m A\partial^m\bar{A}-\frac{1}{4}F_{mn}F^{mn}-\frac{1}{2}J''\partial_mC\partial^mC-\frac{1}{2}J''B_mB^m-\mathcal{V}~,
\end{equation}
with the scalar potential
\begin{equation} \label{pot}
\mathcal{V}=\frac{g^2}{2}{J'}^2+e^{K+2J}{}\biggl[
	K^{-1}_{A\bar{A}}(\mathcal{W}_A+K_A\mathcal{W})(\overbar{\mathcal{W}}_{\bar{A}}+K_{\bar{A}}\overbar{\mathcal{W}})-\bigg(3-2\fracmm{{J'}^2}{J''}\bigg)\mathcal{W}\overbar{\mathcal{W}}
	\biggl]~,
\end{equation}
where we have used the supergravity (bosonic) field components defined by~\footnote{The vertical bars denote the leading field components of the superfields at $\theta=\bar{\theta}=0$.}
\begin{gather*}
2\mathcal{E}|=e,\;\;\;\;\mathcal{D}\mathcal{D}(2\mathcal{E})|=4e\overbar{M}~,\\
\mathcal{R}|=-\frac{1}{6}M,\;\;\;\;\mathcal{D}\mathcal{D}\mathcal{R}|=-\frac{1}{3}R+\frac{4}{9}M\overbar{M}+\frac{2}{9}b_mb^m-\frac{2}{3}i\mathcal{D}_mb^m~,
\end{gather*}
in terms of the vierbein determinant $e\equiv\text{det} e_m^a$, the spacetime scalar curvature $R$, and the minimal set of the supergravity auxiliary fields given by the complex scalar $M$ and the real vector $b_m$. The matter (bosonic) field components  are defined by 
\begin{gather*}
\Phi|=A~,\;\;\;\;\mathcal{D}_\alpha\mathcal{D}_\beta\Phi|=-2\varepsilon_{\alpha\beta} F~,\;\;\;\;\overbar{\mathcal{D}}_{\dot{\alpha}}\mathcal{D}_{\alpha}\Phi|=-2i{\sigma_{\alpha\dot{\alpha}}}^m\partial_mA~,
\\\overbar{\mathcal{D}}\overbar{\mathcal{D}}\mathcal{D}\mathcal{D}\Phi|=16\Box A+\frac{32}{3}ib_a\partial^aA+\frac{32}{3}FM~,
\\V|=C~,\;\;\;\;\mathcal{D}_\alpha\mathcal{D}_\beta V|=\varepsilon_{\alpha\beta} X~,\;\;\;\;\overbar{\mathcal{D}}_{\dot{\alpha}}\mathcal{D}_{\alpha}V|={\sigma_{\alpha\dot{\alpha}}}^m(B_m-i\partial_mC)~,
\\\mathcal{D}_\alpha W^{\beta}|\equiv -\frac{1}{4}\mathcal{D}_\alpha(\overbar{\mathcal{D}}\overbar{\mathcal{D}}-8\mathcal{R})\mathcal{D}^{\beta}V=\frac{1}{2}{\sigma_{\alpha\dot{\alpha}}}^m\overbar{\sigma}^{\dot{\alpha}\beta n}(\mathcal{D}_m\partial_n C+iF_{mn})+{\delta_\alpha}^\beta (D+\frac{1}{2}\Box C)~,
\\ \overbar{\mathcal{D}}\overbar{\mathcal{D}}\mathcal{D}\mathcal{D}V|=\frac{16}{3}b^m(B_m-i\partial_mC)+8\Box C-\frac{16}{3}MX+8D~,
\end{gather*}
in terms of the physical fields ($A$, $C$, $B_m$), the auxiliary fields ($F$, $X$, $D$) and the vector field strength 
$F_{mn}=\mathcal{D}_mB_n-\mathcal{D}_nB_m$. 

When the function $J$ is {\it linear} with respect to its argument (i.e. in the case of the {\it minimal} coupling of the vector multiplet to supergravity), our results agree with the textbook   \cite{Wess:1992cp}.
In the absence of chiral matter, $\Phi=0$, our results also agree with Refs.~\cite{Ferrara:2013rsa,VanProeyen:1979ks}.~\footnote{Our $J$-function and the $C$-function of the inflaton field $\phi$ to be introduced below, differ by their signs from those used in Ref.~\cite{Ferrara:2013rsa,VanProeyen:1979ks}.}

As is clear from Eq.~\eqref{complag}, the absence of ghosts requires $J''(C)>0$, where the primes denote the differentiations with respect to the given argument. In this paper, we restrict ourselves to the K\"ahler potential and the superpotential of the {\it Polonyi model} \cite{Polonyi:1977pj}:
\begin{equation} \label{polonyi}
K= \Phi\overbar{\Phi}~,\qquad \mathcal{W}=\mu(\Phi +\beta)~,
\end{equation}
with the parameters $\mu$ and $\beta$. Unlike Ref.~\cite{Antoniadis:2014oya}, we do {\it not} impose the nilpotency condition $\Phi^2=0$, in order to keep manifest (linear) supersymmetry of our original construction \eqref{hsslag} and to avoid a concern about loosing unitary with the nilpotent superfields at high energies.

Then, on the one side, our model includes the single-field $(C)$ inflationary model, whose $D$-type scalar potential is  given by 
\begin{equation} \label{scalpot}
V(C)=\frac{g^2}{2}({J'})^2
\end{equation}
in terms of an {\it arbitrary} function $J(C)$, with the real inflaton field $C$ belonging to a massive vector supermultiplet.
On the other hand, the Minkowski vacuum conditions (after inflation) can be easily satisfied when $J'=0$, which implies \cite{Polonyi:1977pj}
 \begin{equation} \label{vevs}
 \VEV{A}=\sqrt{3}-1 \qquad {\rm and} \qquad \beta=2-\sqrt{3}~~.
 \end{equation}
This solution describes a {\it stable} Minkowski vacuum with spontaneous SUSY breaking at {\it arbitrary} scale 
$\langle F\rangle=\mu$. The related gravitino mass is given by 
\begin{equation} \label{gmass}
m_{3/2}=\mu e^{2-\sqrt{3}+\VEV{J}}~~.
\end{equation}
There is also a massive (Polonyi) scalar of mass $M_A=2\mu e^{2-\sqrt{3}}$ and a massless fermion in the physical spectrum. 

As regards early Universe phenomenology, our specific model of {\it Polonyi-Starobinsky} (PS) supergravity has the following theoretically appealing features:
\begin{itemize}
\item there is no need to "stabilize" the single-field inflationary trajectory against scalar superpartners
of inflaton, because our inflaton is the only real scalar in a massive vector multiplet,
\item any values of CMB observables $n_s$ and $r$ are possible by choosing the $J$-function,
\item a spontaneous SUSY breaking after inflation takes place at arbitrary scale $\mu$,
\item there are only {\it a few} parameters relevant for inflation and SUSY breaking: the coupling constant $g$ defining the inflaton mass, $g\sim m_{\rm inf.}$, the coupling constant $\mu$ defining the scale of SUSY breaking, $\mu\sim m_{3/2}$, and the parameter $\beta$ in the constant term of the superpotential. Actually,  the inflaton mass is constrained by CMB observations as $m_{\it inf.}\sim 
{\cal O}(10^{-6})$, while $\beta$ is fixed by the vacuum  solution,  so that we have only {\it one} free parameter $\mu$
defining the scale of SUSY breaking in our model (before studying reheating and phenomenology).
\end{itemize}

The (inflaton) scalar potential associated with the Starobinsky inflationary model of $(R+R^2)$ gravity arises when 
\cite{Ferrara:2013rsa}
\begin{equation} \label{starj}
 J(C)=  \frac{3}{2} \left( C- \ln C\right)  
 \end{equation}
that implies
\begin{equation} \label{starjder}
 J'(C) =   \frac{3}{2}\left(1- C^{-1}\right)  \qquad {\rm and} \qquad  J'' (C)=   \frac{3}{2}\left(C^{-2}\right)>0~.
 \end{equation}
According to  \eqref{complag}, a canonical inflaton field $\phi$ (with the canonical kinetic term) is related to the field $C$ by the field redefinition
\begin{equation} \label{canonf}
C =  \exp\left( \sqrt{2/3} \phi\right)~.
\end{equation}
Therefore, we arrive at the (Starobinsky) scalar potential 
\begin{equation} \label{starpot}
V_{\rm Star.}(\phi) = \frac{9g^2}{8}  \left( 1- e^{-\sqrt{2/3}\phi}   \right)^2\quad {\rm with} \quad
m^2_{\it inf.}=9g^2/2~~. 
\end{equation}

The full action \eqref{hsslag} of this PS supergravity in curved superspace can be transformed into a supergravity extension of the $(R+R^2)$ gravity action  by using the (inverse) duality procedure described in Ref.~\cite{Ferrara:2013rsa}.  However, the dual supergravity model is described by a  complicated {\it higher-derivative}  field theory that is inconvenient for studying particle production. Actually, there is also the F-type scalar potential in PS supergravity due to mixing of inflaton and Polonyi scalars, that leads to instability of the Starobinsky inflation described by the D-term alone. However, after adding a Fayet-Iliopoulos (FI)  term \cite{Fayet:1974jb} together with its supersymmetric completion \cite{Cribiori:2017laj} to the Lagrangian (\ref{hsslag}) and modifying the  $J$-function above, the Starobinsky inflation can be restored, and the inflaton-Polonyi mixing can be suppressed \cite{Aldabergenov:2017hvp}. The FI term does not affect the phenomenology discussed in this paper, as long as $\VEV{J}$ is negative and close to zero \cite{Aldabergenov:2017hvp}, as we always assume.

Another nice feature of our model is that it can be rewritten as a supersymmetric (abelian and non-minimal) gauge theory coupled to supergravity in the presence of a {\it Higgs} superfield $H$, resulting in the super-Higgs effect with simultaneous spontaneous breaking of the gauge symmetry and supersymmetry. Indeed, the 
$U(1)$ gauge symmetry of the original Lagrangian (\ref{hsslag}) allows us to choose a different ({\it Wess-Zumino}) 
supersymmertic gauge by "gauging away" the chiral and anti-chiral parts of the general superfield $V$ via the appropriate choice of the superfield parameters $Z$ and $\overbar{Z}$ as
\begin{gather*}\label{wzgauge}
V|=\mathcal{D}_\alpha\mathcal{D}_\beta V|=\overbar{\mathcal{D}}_{\dot{\alpha}}\overbar{\mathcal{D}}_{\dot{\beta}} V|=0,\\
\overbar{\mathcal{D}}_{\dot{\alpha}}\mathcal{D}_{\alpha}V|={\sigma_{\alpha\dot{\alpha}}}^m B_m~,
\\
\mathcal{D}_\alpha W^{\beta}|=\frac{1}{4}{\sigma_{\alpha\dot{\alpha}}}^m\overbar{\sigma}^{\dot{\alpha}\beta n}(2iF_{mn})+{\delta_\alpha}^\beta D~,
\\ \overbar{\mathcal{D}}\overbar{\mathcal{D}}\mathcal{D}\mathcal{D}V|=\frac{16}{3}b^mB_m+8D~.
\end{gather*} 
Then the bosonic part of the Lagrangian in terms of the superfield components in Einstein frame, after elimination of the auxiliary fields and Weyl rescaling, reads \cite{Aldabergenov:2017bjt}
\begin{multline} \label{dhlag2}
e^{-1}\mathcal{L}=-\frac{1}{2}R-K_{AA^*}\partial^m A\partial_m\bar{A}-\frac{1}{4}F_{mn}F^{mn}-2J_{h\bar{h}}\partial_mh\partial^m\bar{h}-\frac{1}{2}J_{V^2}B_mB^m\\+iB_m(J_{Vh}\partial^mh-J_{V\bar{h}}\partial^m\bar{h})-\mathcal{V}~,
\end{multline}
where $h$, $\bar{h}$ are the Higgs field and its conjugate.

The standard $U(1)$ Higgs mechanism arises with the canonical function $J=\frac{1}{2}he^{2V}\bar{h}$, where we
have chosen $g=1$ for simplicity.  As regards the Higgs sector, it leads to 
\begin{equation}
e^{-1}\mathcal{L}_{Higgs}=-\partial_mh\partial^m\bar{h}+iB_m(\bar{h}\partial^mh-h\partial^m\bar{h})-h\bar{h}B_mB^m-\mathcal{V}~.
\end{equation}
After changing the variables $h$ and $\bar{h}$ as
\begin{equation}
h=\frac{1}{\sqrt{2}}(\rho+\nu)e^{i\zeta},\;\;\;\bar{h}=\frac{1}{\sqrt{2}}(\rho+\nu)e^{-i\zeta}~,\label{paramh}
\end{equation}
where $\rho$ is the (real) Higgs boson, $\nu\equiv \langle h\rangle=\langle \bar{h}\rangle$ is the Higgs VEV, and $\zeta$ is the Goldstone boson,  the unitary gauge fixing of $h\rightarrow h'=e^{-i\zeta}h$ and $B_m\rightarrow B'_m=B_m+\partial_m\zeta$, leads to the standard result \cite{Weinberg:1973ew}
\begin{equation}
e^{-1}\mathcal{L}_{Higgs}=-\frac{1}{2}\partial_m\rho\partial^m\rho-\frac{1}{2}(\rho+\nu)^2B_mB^m-\mathcal{V}~.
\end{equation}

The same result is also achieved by considering the super-Higgs mechanism where, in order to get rid of the Goldstone mode, one uses the super-gauge transformations \eqref{supergh} and \eqref{supergv}, and defines the relevant field components of $Z$ and $i(Z-\overbar{Z})$ as
\begin{equation}
Z|=\zeta+i\xi~,\;\;\; \frac{i}{2}\overbar{\mathcal{D}}_{\dot{\alpha}}\mathcal{D}_{\alpha}(Z-\overbar{Z})|=\sigma_{\alpha\dot{\alpha}}^m\partial_m\zeta~.
\end{equation}
Examining the lowest components of the transformation \eqref{supergh}, one can easily see that the real part of $Z|$ cancels the Goldstone mode of \eqref{paramh}. Similarly, when applying the derivatives $\overbar{\mathcal{D}}_{\dot{\alpha}}$ and  $\mathcal{D}_{\alpha}$ to \eqref{supergv} and taking their lowest components (then $\overbar{\mathcal{D}}_{\dot{\alpha}}\mathcal{D}_{\alpha}V|=\sigma^m_{\alpha\dot{\alpha}}B_m$), one finds that the vector field "eats up" the Goldstone mode as
\begin{equation}
B'_m=B_m+\partial_m\zeta~.
\end{equation}

The Minkowski vacuum after inflation can be easily lifted to a {\it de Sitter} vacuum (Dark Energy) in our model by the simple modification of the Polonyi sector and its parameters as \cite{Aldabergenov:2017bjt}
\begin{equation}
\VEV{A}=(\sqrt{3}-1)+\fracmm{3-2\sqrt{3}}{3(\sqrt{3}-1)}\delta+\mathcal{O}(\delta^2)~,\quad \beta=(2-\sqrt{3})+\fracmm{\sqrt{3}-3}{6(\sqrt{3}-1)}\delta+\mathcal{O}(\delta^2)~.
\end{equation}
It leads to a small positive cosmological constant
\begin{equation}
V_0=\mu^2e^{\alpha^2}\delta=m^2_{3/2}\delta\label{dsvac}
\end{equation}
and the superpotential  VEV
\begin{equation}
\langle \mathcal{W}\rangle=\mu(\VEV{A} +\beta)=\mu(a+b-\frac{1}{2}\delta)~,
\end{equation}
where $a\equiv(\sqrt{3}-1)$ and $b\equiv(2-\sqrt{3})$ is the SUSY breaking vacuum solution to the Polonyi parameters in the absence of a cosmological constant (see Ref.~\cite{Linde:2016bcz} also).

\section{Gravitino and Polonyi production} 

\noindent 

In this Section, we consider the gravitino ($\psi_{\mu}$) and Polonyi  ($A$) particles 
production during inflation and after it.  
We assume that  all other (heavy) SUSY particles (not present in our model) 
have masses larger than those  of Polonyi and gravitino (High-scale SUSY), with gravitino as the LSP (the lightest superpartner of known particles) and as the cold  Dark Matter (SHGDM).

There are several competitive sources of particle production in our model. First,
gravitino and Polonyi particles can be  produced via Schwinger's effect  (out of vacuum) sourced by inflation. 
Since the mass of a Polonyi particle is higher than two gravitino masses (Sec.~2), the former is unstable, and
 decays into two gravitinos, $A\rightarrow \psi_{3/2}\psi_{3/2}$. Second, both gravitino and Polonyi can be produced by inflaton decays, during oscillations of the inflaton field around its minimum after inflation.
A competition between  the gravitino/Polonyi creation during inflation and their production by inflaton decays  is known to be very sensitive to the mass hierarchy. It is, therefore, very instructive to
study them in our model (Sec.~2) that is minimalistic and highly constrained. 

Since the exact equations of motion in our model (Sec.~2) are very complicated, in this Section we take them only in the leading order with respect to the inverse Planck mass. Let us begin with Polonyi and gravitino production  during inflation by ignoring for a moment their couplings to inflaton. The effective action  of the Polonyi field in the FLRW background reads 
\begin{equation}
\label{Sdt}
I[A]=\int dt \int d^{3}x \fracmm{a^{3}}{2}\left(\dot{A}^{2}-\fracmm{1}{a^{2}}(\nabla A)^{2}-M_{A}^{2}A^{2}-\zeta R A^{2}\right)~,
\end{equation}
where the non-minimal coupling  constant (of Polonyi field  to gravity) is $\zeta=1$ in our case (Sec.~2),
$A$ is the Polonyi field, $a$ is the FLRW scale factor, 
$M_{A}$ is the Polonyi mass, and $R$ is the Ricci scalar. 

A mode expansion of the Polonyi field in terms of the conformal time coordinate $\eta$ reads

\begin{equation}
\label{XX}
A({\bf x})=\int d^{3}k (2\pi)^{-3/2}a^{-1}(\eta)\left[ b_{k}h_{k}(\eta)e^{i{\bf k}\cdot {\bf x}}+b_{k}^{\dagger}h^{*}_{k}(\eta)e^{-i{\bf k}\cdot {\bf x}}\right]\, ,
\end{equation}
where $b,b^{\dagger}$  are the (standard)  creation/annihilation operators, and the 
coefficient functions $h,h^{+}$ are properly normalized as 
\begin{equation}\label{hh}
h_{k}h'^{*}_{k}-h'_{k}h^{*}_{k}=i~~.
\end{equation}

It follows from Eqs.~(\ref{Sdt}) and (\ref{XX}) that the  equations of motion of the modes are given by
\begin{equation}
\label{modes}
h''_{k}(\eta)+\omega_{k}^{2}(\eta)h_{k}(\eta)=0~, \quad { \rm where}  \quad \omega_{k}^{2}=k^{2}+M_{A}^{2}a^{2}+5\fracmm{a''}{a}~~,
\end{equation}
and we have defined $h''=d^{2} h/d\eta^{2}$ with respect to the conformal time  $\eta$.  For our purposes, it is convenient to rescale Eq.~(\ref{modes})  by  some reference constants  $a(\eta_{*})\equiv a_*$ and  
$H(\eta_{*})=H_{*}$ to be specified later, and rewrite it as
\begin{equation}
\label{res}
h''_{\tilde{k}}(\tilde{\eta})+(\tilde{k}^{2}+b^{2}\tilde{a}^{2})h_{\tilde{k}}(\tilde{\eta})=0\, ,
\end{equation}in terms of the rescaled quantities
$$\tilde{\eta}=\eta a_{*}H_{*}~,\quad \tilde{a}=a/a_{*}~,\quad \tilde{k}=k/(H_{*}a_{*})~~.$$

Similarly, the gravitino field is governed by the  massive  Rarita-Schwinger action 
\begin{equation}
\label{Rarita}
I[\psi]=\int d^{4}x \,e\, \bar{\psi}_{\sigma}\mathcal{R}^{\sigma}\{\psi\}\, ,
\end{equation}
in terms of the gravitino field strength 
\begin{equation}
\label{RC}
\mathcal{R}^{\sigma}\{\psi\}=i \gamma^{\sigma\nu\rho}\mathcal{D}_{\nu}\psi_{\rho}+m_{3/2}\gamma^{\sigma\nu}\psi_{\nu}
\end{equation}
and the supercovariant derivative
\begin{equation}
\label{cov}
\mathcal{D}_{\mu}\psi_{\nu}=\partial_{\mu}\psi_{\nu}+\frac{1}{4}\omega_{\mu ab}\gamma^{ab}\psi_{\nu}-\Gamma_{\mu\nu}^{\rho}\psi_{\rho}~~,
\end{equation}by using the notation
$\gamma^{\mu_{1}...\mu_{n}}=\gamma^{[\mu_{1}}....\gamma^{\mu_{n}]}$ with unit weight of the antisymmetrization. The supergravity torsion  is of the second order with respect to the inverse Planck mass, so that it can be ignored. The $\Gamma_{\mu\nu}^{\rho}$ can be represented by  the symmetric Christoffel symbols,  but  they are cancelled from the action (\ref{Rarita}).

The gravitino equation of motion now reads 
\begin{equation}
\label{eqRSS}
(i  \mathcal{\slashed{D}}-m_{3/2})\psi_{\mu}-\left(i\mathcal{D}_{\mu}+\frac{m_{3/2}}{2}\gamma_{\mu}\right)\gamma \cdot \psi=0~~.
\end{equation}
In the flat FLRW background,
Eq.~(\ref{eqRSS}) becomes  
\begin{equation}
\label{psiEOMS}
i\gamma^{mn}\partial_{m}\psi_{n}=-\left(m_{3/2}+i\fracmm{a'}{a}\gamma^{0}\right)\gamma^{m}\partial_{m}\psi~~,
\end{equation}
where we have
\begin{equation}
\label{cond}
e_{\mu}^{a}=a(\eta) \delta_{\mu}^{a}~,\quad m_{3/2}=m_{3/2}(\eta)~,\quad \omega_{\mu ab}=2\dot{a}a^{-1}e_{\mu[a}e_{b]}^{0}~~.
\end{equation}
A solution to Eq.~(\ref{psiEOMS}) for the helicity $3/2$ modes 
 reads 
\begin{equation}
\label{grav}
\psi_{\mu}(x)=\int d^{3}{\bf p}(2\pi)^{-3}(2p_{0})^{-1}\sum_{\lambda}\{ e^{i{\bf k}\cdot {\bf x}}b_{\mu}(\eta,\lambda)a_{k\lambda}(\eta)+e^{-i{\bf k}\cdot {\bf x}}b_{\mu}^{C}(\eta,\lambda)a_{k\lambda}^{\dagger}(\eta) \}~~.
\ee

We find that the equations of motion for the $3/2$-helicity gravitino modes have the same form as that of 
Eq.~(\ref{modes}), namely,
\begin{equation}
\label{Modes1}
b''_{\mu}(\eta,\lambda)+\hat{C}(k,a)b'_{\mu}(\eta,\lambda)+\omega^{2}(k,a)b_{\mu}(\eta,\lambda)=0~,
\end{equation}
\begin{equation}
\label{Ca}
\hat{C}(k,a)b'_{\mu}(\eta,\lambda)=-2i\gamma^{\nu i}k_{i}\gamma_{\nu\eta}\partial^{\eta}b_{\mu}-2\gamma_{\nu}(m_{3/2}+i\frac{a'}{a}\gamma^{0})i\gamma^{\nu\eta}\partial_{\eta}b_{\mu}~~,
\end{equation}
\begin{equation}
\label{omega2}
\omega^{2}(k,a)/2= k^{2}+m_{3/2}^{2}+2i\frac{a'}{a}\gamma^{0}m_{3/2}
-\left(\frac{a'}{a}\right)^{2}~~.
\end{equation}

It is customary to write them down (similarly to the well known
relation between Dirac and Klein-Gordon equations) as
\begin{equation}
\label{PP}
P_{\nu}P^{\nu}b_{\mu}(\eta,\lambda)=0~~,
\end{equation}
where, in our case, we have
\begin{equation}
\label{PPP}
P^{\nu}=i\gamma^{\nu\eta}\partial_{\eta}-\gamma^{\nu i}k_{i}-\left(m_{3/2}+i\frac{a'}{a}\gamma^{0} \right)\gamma^{\nu}=0\, . 
\end{equation}
Equation (\ref{Modes1}) can be rescaled in the same way as  Eq.~(\ref{res}). 

Interactions of gravitino with matter fields can be described in terms of the {\it effective} gravitino mass $M_{3/2}$ that is a function of matter fields in the Rarita-Schwinger equation, with $m_{3/2}=\VEV{M_{3/2}}$.
In our model with the matter given by inflaton and Polonyi scalars (Sec.~2),  we find
\begin{equation} \label{effmg}
M_{3/2}(\phi, \tilde{A})=\mu M_{Pl}^{-1}\exp\left[(1/\sqrt{6})M_{Pl}^{-1}\phi+M_{Pl}^{-2}(\bar{\tilde{A}}\tilde{A}+\alpha\bar{\tilde{A}}+\alpha\tilde{A}+\alpha^2)\right](\tilde{A}+\alpha+\beta)~,
\end{equation}
where $\alpha\equiv\VEV{A}$ and $A=\alpha+\tilde{A}$, in terms of  inflaton field $\phi$ and Polonyi scalar $\tilde{A}$ with $\VEV{\tilde{A}}=0$, and we have restored the dependence upon Planck mass.

In order to obtain the number density of produced particles, we perform a Bogoliubov 
transformation,
\begin{equation}
\label{BOGO}
h_{k}^{\eta_{1}}(\eta)=\alpha_{k}h_{k}^{\eta_{0}}(\eta)+\beta_{k}h_{k}^{*\eta_{0}}(\eta)\, .
\end{equation}
This transformation is supposed to be done from the vacuum solution 
with the boundary condition
$\eta=\eta_{in}$, corresponding to the initial time 
of inflation, to the final time $\eta=\eta_{f}$  when  particles 
are no longer created from inflation. 
Since we have
$a'/a^{2}\ll 1$ and
 $ba/k\ll 1$,  we can take the extremes as
$\eta_{in}=-\infty$ and $\eta_{f}=+\infty$
in the semiclassical approximation. 
Given  such boundary conditions, 
the energy density of Polonyi particles produced 
during inflation is given by
\begin{equation}
\label{S}
\rho_{A}(\eta_{})=M_{A}n_{A}(\eta_{})=M_{A}H_{inf}^{3}\left(\fracmm{1}{\tilde{a}(\eta_{})} \right)^{3}\mathcal{P}_{A}~~ ,
\end{equation}
where we have used the standard notation
\begin{equation}
\label{power}
\mathcal{P}_{A}=\fracmm{1}{2\pi^{2}}\int_{0}^{\infty}d\tilde{k}\tilde{k}^{2}|\beta_{\tilde{k}}|^{2}~~. 
\end{equation}
Similar equations are valid for gravitino, with the power spectrum
\begin{equation}
\label{power}
\mathcal{P}_{\psi}=\fracmm{1}{2\pi^{2}}\int_{0}^{\infty}d\tilde{k}\tilde{k}^{2}|b_{\mu}b^{C\mu}|~~. 
\end{equation}

\begin{figure}[t!!]
\begin{center}
\vspace{1cm}
\includegraphics[width=15cm,height=8cm,angle=0]{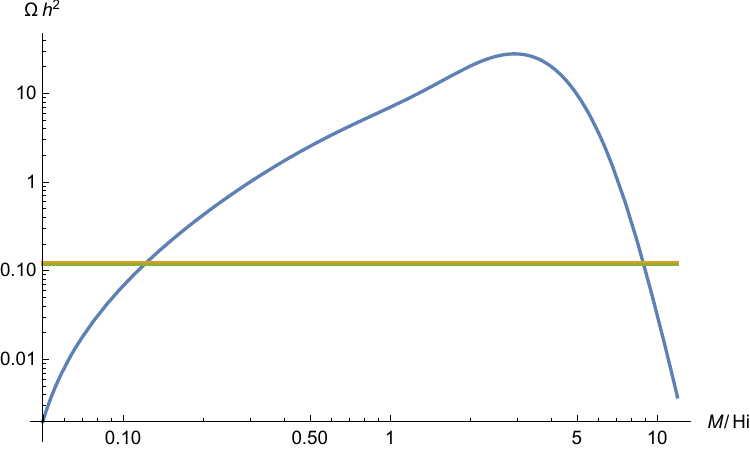}
\caption{ Numerical simulations of the produced gravitino
mass density (normalized) as a function of the Polonyi mass
parameter are displayed in blue, in the parameter range
compatible with inflation, reheating and leptogenesis (at the reference point
$N_{e}=55$): $n_{s}=0.964$, $r=0.004$, $m_{inf}=3.2\cdot 10^{13}\, {\rm GeV}$,
$H_{inf}=1.4\cdot 10^{14}\, {\rm GeV}$,  and $T_{reh}=3\cdot 10^{9}\, {\rm GeV}$. 
The right amount of cold DM: $\Omega_{3/2}h^2=\Omega_{DM}h^2=0.11$ (in orange) is generated 
when the Polonyi mass is  $M_A\approx 2m_{3/2}= (1.54\pm 0.2)\times 10^{13}\, {\rm GeV}$ that is compatible  
with the Polonyi mass inferred from  Starobinsky inflation and reheating in Eq.~(\ref{GM}) below.} \label{fig:1}
\end{center}
\end{figure}

Some comments are in order.

(i) Technical details about the power spectrum and our estimate of the normalized value of $\mathcal{P}_{A}$ to be 
of the order  $\exp\left[-O(1)M_{A}/H_{inf}\right]$ are given in Appendix.

(ii) Our proposal about Polonyi particles produced during inflation reads
\begin{equation}
\label{rewe}
\Omega_{A}h^{2}\simeq \Omega_{R}h^{2}\left(\fracmm{T_{\rm reh}}{T_{0}} \right)\fracmm{8\pi}{3}\left( \fracmm{M_{A} }{M_{Pl}}\right)\fracmm{n_{A}(t_{f})}{M_{Pl}H^{2}(t_{f})}~~,
\end{equation}
where $M_{A}$ is the Polonyi mass, $\Omega_{R}h^2\simeq 4.31 \times 10^{-5}$ is the fraction 
of critical energy density that is in 
radiation today, $\Omega_{A}h^2$ is 
fraction of the critical energy density of produced Polonyi fields (and a similar estimate for gravitino). 
 Though we do not have a rigorous proof of Eq.~(\ref{rewe}), it can be argued  by starting from
 \begin{equation}
\label{RHOOO}
\fracmm{\rho_{A}(t_{0})}{\rho_{R}(t_{0})}=\fracmm{\rho_{A}(t_{\rm reh})}{\rho_{R}(t_{\rm reh})}\left(\fracmm{T_{\rm reh}}{T_{0}}\right),
\end{equation}
  where $\rho_{R}$ is the energy density of radiation,  $\rho_{A}$ is the Polonyi energy density, 
 $T_{\rm reh}$ and $T_{0}$ are the temperature of the Universe at reheating time $t_{\rm reh}$ and today $t_{0}$, respectively.  Assuming that Polonyi particles are produced after the de Sitter phase 
 $t_{e}$, when the transition to the coherent oscillation phase begins, the inflaton and Polonyi energy densities will be redshifted  with almost the same rate. This scaling holds until the reheating stage  finishes and the radiation dominated epoch begins.  Assuming that most of the energy density is  converted into radiation contribution with
 \begin{equation}
 \rho_{R}\simeq \rho_{c}=\fracmm{3H^{2}M_{Pl}^{2}}{8\pi}\, ,
 \end{equation}
 we obtain 
 \begin{equation}
 \label{reply}
 \fracmm{\rho_{A}(t_{\rm reh})}{\rho_{R}(t_{\rm reh})}\simeq \fracmm{8\pi}{3}
 \fracmm{\rho_{A}(t_{e})}{M_{Pl}^{2}H^{2}(t_{e})}~ ,
 \end{equation}
 where $H(t_{e})$ is the Hubble parameter at a fixed time $t=t_{e}$. Then Eq.~(\ref{rewe}) follows from Eq.~(\ref{reply}).
 
(iii)  According to Eq.~(\ref{S}), relating Hubble scale, Polonyi mass and the {\it desiderata} Polonyi energy-density, there is about $8th$-orders-of-magnitude suppression of the energy-density.  According to (i), the normalized power spectrum $\mathcal{P}_{A}$ cannot provide such suppression 
with our values for $M_A$ and $H_{inf}$. However, it comes from the dilution factor $(\tilde{a})^{-3}=(a_{f}/a_{i})^{-3}$ in Eq.~(\ref{S}). 

Our semi-analytical estimations for Eq.~(\ref{S}) indicate that almost all Polonyi particles are produced in an excursion of the inflaton field around $\phi_{e}\equiv\phi(t_{e})$  with $\Delta \phi\simeq 0.2$. The value of the dilution factor can be estimated from 
\begin{equation}
\label{int}
(a(t_{f})/a(t_{i}))^{-3}=\exp\left[-24\pi \int_{\phi_{f}}^{\phi_{i}}d\phi\, V^{-1}(\phi)V_{,\phi}(\phi)\right]=\exp(-\Delta \Phi)~, 
\end{equation}
where we have defined $\Delta \Phi=\Phi(\phi(t_{i}))-\Phi(\phi(t_{f}))$, having in mind that $\phi(t_{i})>\phi(t_{f})$ and
\begin{equation}
\label{DeltaPhi}
\Phi(\phi)=48\pi \sqrt{2/3} e^{-\sqrt{2\phi/3}}(1-e^{-\sqrt{2\phi/3}})^{-1}\, .
\end{equation}
After integrating over the effective particle production region $\Delta \phi$, we find 
$\Delta \Phi=18.2$, i.e. 
\begin{equation}
\label{get}
(a(t_{f})/a(t_{i}))^{-3}\simeq \exp(-18.2)\simeq 10^{-8}~,
\end{equation}
leading to the correct CDM amount.

(iv) To get the masses $M_A$ and $m_{3/2}\equiv m_{\psi}$ in a different way, we have to add a few more assumptions about details of reheating. Since our SHGDM scenario is based on Starobinsky inflation, all cosmological parameters can be  fixed modulo the e-foldings number $N_{e}$ that is between $50$ and $60$ for compatibility with CMB observations. This also allows us to estimate the error margin for the masses in question at about 20\%.

Let us take $N_{e}=55$ as the best fitting (reference) point \cite{Ketov:2012yz,Ellis:2015pla}.  This leads to 
the following set of the cosmological (inflation) parameters \cite{Gurovich:1979xg}:
\begin{equation} \label{cosmdata}
n_{s}=0.964,\quad r=0.004,\quad m_{inf}=3.2\cdot 10^{13}\, {\rm GeV},\quad H_{inf}=\p M_P\sqrt{P_{g}/2}=
1.4\cdot 10^{14}\, {\rm GeV}~,
\end{equation}
where $M_{P}=2.44\cdot 10^{18}~{\rm GeV}$ is the reduced Planck mass, and $P_g$ stands for tensor perturbations.
In our SHGDM scenario, well below the inflaton mass scale, the low-energy theory is given by the Standard Model (SM) that has the effective number of d.o.f. as $g_{*}=106.75$. Then, it is reasonable to assume that all the SM particles were generated by perturbative inflaton decay, whose reheating temperature is well known in Starobinsky inflation \cite{star1982,Vilenkin:1985md,Ketov:2012yz},
\begin{equation}
\label{T1}
T_{reh}=\left( \fracmm{90}{\pi^{2}g_{*}}\right)^{1/4}\sqrt{\Gamma_{tot}M_{P}}=3\cdot 10^{9}\,{\rm GeV} \, . 
\end{equation}
This value is also consistent with the successful leptogenesis scenario of Ref.~\cite{Antusch:2006gy} based on the see-saw type-I mechanism, that requires the reheating temperature to be higher than $1.4\times 10^{9}~{\rm GeV}$. 

On the other hand, the reheating temperature for heavy gravitino is given by \cite{Jeong:2012en,Domcke:2017rzu}
\begin{equation}
\label{T2}
T_{reh}=1.5\cdot 10^{8}~{\rm GeV}\left(\fracmm{80}{g_{*}} \right)^{1/4}\left(\fracmm{m_{3/2}}{10^{12}\,{\rm GeV}} \right)^{3/2}\, . 
\end{equation}
Combining Eqs.~(\ref{T1}) and (\ref{T2}), we arrive at the gravitino and Polonyi masses
\begin{equation}
\label{GM} 
m_{3/2}=(7.7\pm 0.8)\cdot 10^{12}\, {\rm GeV} \quad {\rm and} \quad M_{A}= 2e^{-\VEV{J}}m_{3/2}\approx 2m_{3/2}~.
\end{equation}
These masses are compatible with the correct abundance of cold DM, according to our numerical estimates in Fig.~1,
that lends further support towards our conjecture in Eq.~(\ref{rewe}).

(v) The decay rate of Polonyi particle into two gravitinos is given by \cite{Endo:2007sz,Nakayama:2014xca}
\begin{equation}
\label{GAMMA}
\Gamma(A\rightarrow \psi_{3/2}\psi_{3/2})\simeq \fracmm{3}{288\pi}M_{A}^{3}/m_{3/2}^{2}\simeq 2.6\times 10^{-2}m_{3/2}~~.
\end{equation}
This channel is a direct consequence of the gravitino mass generation mechanism from the non-vanishing Polonyi vacuum expectation value. In addition, it implies that Polonyi particles rapidly decay into gravitinos. Moreover, since we
have $m_{3/2}=7.7\cdot 10^{12}\, {\rm GeV}$, it implies   $\Gamma\ll H_{inf}$.
This means that the decay time scale  is much larger then the production time 
during inflation, i.e. $\tau_{A\rightarrow \psi_{3/2}\psi_{3/2}}\gg \tau_{inflation}$.
As a consequence, the decays of Polonyi particles
into gravitinos are subleading and negligible during inflation. 
Therefore, gravitino and Polonyi particles  are 
independently created 
during inflation.  After the reheating, 
Polonyi particles will completely decay into gravitinos (see below). 
In particular, the Polonyi number density  $n_{S}$ gets
 transformed into a contribution to the gravitino number density
 $\Delta n_{\Psi}=2n_{A}$. Since the Polonyi  mass is about two times of the gravitino mass, 
the Polonyi energy density before its decays is 
completely converted into gravitinos, 
i.e. $(\Delta \Omega_{\psi}) h^{2}=\Omega_{S} h^{2}$. 

In Fig.~1 we show a numerical simulation of the produced gravitino
mass density as a function of the Polonyi mass.~\footnote{We chose the lower (on the left) intersection
point in Fig.~1 because the higher (on the right) intersection point leads to the heavy Polonyi particle becoming a spectator during inflation and reheating that is inconsistent with our approach.}

The spectrum is composed of two contributions: (a) the Polonyi energy-density spectrum produced during inflation,  converted into gravitinos after reheating; and (b) the energy-density spectrum of gravitinos produced during inflation. The first contribution largely dominates over the second one. 
Intriguingly, the Polonyi and gravitino masses inferred from inflation, reheating and leptogenesis bounds 
are well compatible with the correct amount of CDM.

Next, let us consider the gravitino and Polonyi production from inflaton decays. As regards gravitino, its coupling to inflaton arises  from  the Weyl rescaling of  vierbein in the gravitino action. The gravitino kinetic term does not contribute because of conformal flatness of the FLRW universe, so that the only source of gravitino production is given by the gravitino mass term~\footnote{Similarly, we can ignore the massless fermion present in the spectrum of our model because  the expansion of (conformally flat) FLRW universe does not lead to perturbative production of massless particles.}
\beq \lb{gramt}
 {\cal L}_{\rm mass} = -\frac{1}{2}e^{G_{\rm tot}/2}\bar{\j}_{\m}\g^{\m\n}\j_{\n} ~~,
 \eeq
where the $G_{\rm tot}$ in our model is given by
\beq \lb{gravm}
G_{\rm tot} = K +\ln \abs{W}^2 + 2J~~,
\eeq
and the gravitino mass is by $m_{3/2}=\VEV{e^{G_{\rm tot}/2}}$.  The mass term in the form \eqref{gramt} also shows the Polonyi and inflaton couplings to gravitino. 

The perturbative decay rate of inflaton $\f$  into a  pair of gravitino is given by \cite{Endo:2006z}
\beq \lb{pergrav}
\G_{\f\to\j_{3/2}\j_{3/2}} = \fracmm{ \abs{G_{\f}}^2}{288\pi}\fracmm{m_{\rm inf}^5}{m^2_{3/2}M^2_{Pl}}~~.
\eeq
In our case, the factor $G_{\f}$ vanishes at the minimum of the inflaton scalar potential \eqref{starpot}, because the inflaton VEV also vanishes. So, the perturbative production of gravitino from inflaton decays is suppressed. One may also wonder about a non-perturbative gravitino production  from inflaton decays, as was studied e.g., in Refs.~\cite{Greene:1998nh,Giudice:1999fb}. However, unlike the usual Yukawa couplings, the coupling of gravitino to inflaton in our model is given by the exponential factor in \eqref{gramt} that never vanishes. Hence, the gravitino production due to inflaton decays  can be ignored in our model.

The situation is different  with the Polonyi production due to inflaton decays. Inflaton is heavier than Polonyi particle by the factor of two approximately, according to Eqs.~(\ref{cosmdata}) and (\ref{GM}).
The perturbative decay rate of inflaton into a pair of Polonyi scalars is (see e.g., Ref.~\cite{Ketov:2012yz} for a review)
\beq \lb{perpol}
\G_{\f\to AA} = \fracmm{1}{192\pi}\fracmm{m_{\rm inf}^3}{M^2_{Pl}}~~.
\eeq
One may expect that the non-perturbative pre-heating in this case can be significantly more efficient due to a broad parametric resonance \cite{Kofman:1994rk} that is a rather generic feature for coupled scalars. Indeed, in our model,  inflaton is mixed with Polonyi field via the scalar potential  \eqref{pot}. 
 An expansion of the scalar potential  \eqref{pot} with the $J$-finction given by Eq.~(\ref{starj}), with respect to both scalar fields  $\f$ and $A$, gives rise to the quartic interacion term 
\beq \lb{quartic}
{\cal L}_{\f\f\to AA} = \l\f^2\bar{A}A~~,
\eeq
whose dimensionless coupling constant $\l$ does not vanish. It implies that the broad  parametric resonance can happen in our model along the lines of Ref.~\cite{Kofman:1994rk}, while it could lead to the enhancement of  the perturbative production of Polonyi particles  up to the factor of ${\cal O}(10^5)$ --- see. e.g., Ref.~\cite{Ketov:2012se} for the example of numerical calculations. However, our direct calculation yields
\beq \label{lambdaval}
\l = 10 e^{7-2\sqrt{3}}\fracmm{\m^2}{M^2_{Pl}}~~,
\eeq
so that the coupling constant $\l$ is of the order ${\cal O}(10^{-7})$ or less in our model, and this value fully compensates any  possible enhancement of the Polonyi  production by the broad parametric resonance. In short, the Polonyi production from inflaton decays is just perturbative in our case.  The Polonyi particles produced in this channel quickly decay into gravitinos, with the perturbative decay rate (\ref{GAMMA}) implying that the gravitino production from inflaton decays is sub-leading with respect to Schwinger's effect shown in Fig.~1. The former channel would be kinematically suppressed if the Polonyi mass  were higher than half of the inflaton mass,  because then the inflaton two-particle decay into two Polonyi particles would be forbidden. 
 
(vi) To the end of this Section, we note that the pre-heating stage with the  dust-like equation of state $p=0$, started at the end of inflation at $t_i \sim 1/H$ and finished in the period of reheating at $t_f \sim M_{Pl}/T_{\rm reh}^2$, is sufficiently long to provide the  growth of density fluctuations and the formation of nonlinear structures of gravitationally bounded systems. In particular, the Primordial Black Holes (PBHs)  may be formed at this stage. Later on, the PBHs can be evaporating and converting about $1/N (1 \rm g/M)$ of their masses to gravitinos, where $N$ is the number of evaporated species with account of their statistical weight, and $M$ is the PBH mass. Here  we have taken into account that at $M \le 1 \rm g$ the temperature of Hawking evaporation is $T_{\rm ev} \ge 10^{13}~{\rm GeV}$, so that the fraction of evaporated gravitino is determined by the ratio of their statistic weight to the statistic weight of all evaporated species. During the pre-heating stage at $t_i \le t \le t_f$, the gravitationally bounded systems are formed in the mass range 
\beq M_o \le M \le M_{max}, \label{PBH}\eeq 
where 
\beq M_o= M_{Pl}^2/H_{inf} \label{Mo} \eeq 
is the mass within the cosmological horizon at the beginning of pre-heating, and 
\beq
M_{max}= \delta^{3/2}M_{Pl}^2 t_f 
\eeq
is the mass of the gravitationally bounded systems formed at the end of pre-heating. 
Here $\delta$ is the amplitude of density fluctuations.

According to Ref.~\cite{Khlopov:1985jw}, the minimal estimation of the probability of the PBH formation at this stage is determined by  consideration of  their direct collapse in the  black holes with the special symmetric and homogeneous gravitationally bound configurations \cite{polnarev1}, and is given by $\sim \delta^{13/2}$ \cite{polnarev2}. For $\delta \approx 10^{-5}$, the estimated fraction of the total density at the end of pre-heating stage $\b$, corresponding to the PBH in the range (\ref{PBH})  is of the order of 
\beq 
\b < 10^{-32}~~.
\eeq
Hence, the contribution of gravitino produced in the evaporation of these black holes is negligible at the beginning of the matter dominated stage with $T_{\rm md}=1~{\rm eV}$ as
\beq
\Omega_g < \fracmm{T_{\rm reh}}{T_{\rm md}}\fracmm{1}{N}\b < 10^{-17}~~,
\eeq
where we have also used $N \sim 10^3$. A formation of black holes in the course of evolution of the gravitationally bounded systems formed during the pre-heating stage can significantly increase the value of $\b$, though addressing the problem of evolution of the gravitationally bounded systems of scalar fields deserves a separate study.

The situation drastically changes when  inflation ends by a {\it first} order phase transition when bubble collisions lead to copious production of black holes at 
$\b \sim 0.1$ with the mass (\ref{Mo}) \cite{miniPBH,miniPBH2}. Evaporation of these PBHs leads to a fraction of the total density by the end of pre-heating of the order $10^{-4}$ in the form of gravitino, and it results in the gravitino dominated stage at $T \sim 10^{-4} T_{\rm reh} \sim 10^5 \rm ~GeV$. This huge gravitino overproduction {\it excludes} a possibility of the first order phase transition by the end of inflation.  This is impossible in the single-field inflation, also by considering an extra axion-like field and extra moduli decoupled during the slow-roll epoch. The inflaton in our model has the characteristic Starobinsky potential that, as is well known,  
does not lead to any violent phase transition after inflation. The Polonyi field does not alter this situation.
However, in a more general case, in which other scalar and pseudo-scalar fields enter the slow-roll dynamics,  these extra fields can have scalar potentials ending in false minima during the reheating. In such case,  the tunnelling from the false minima to the true minima can induce bubbles in the early  Universe, catalising an efficient formation of PBHs. These issues deserve a more detailed investigation  beyond the scope of our paper.

\section{Conclusion and outlook}

In this paper we studied the gravitino production in the context of Starobinsky-Polonyi $\mathcal{N}=1$ supergravity 
with the inflaton field belonging to a massive vector multiplet, and the mass hierarchy  $m_{\rm inf}>2M_A>4m_{3/2}$
close to the bounds (by the order of magnitude).  On the one hand, we found the regions in the parameter space where the
gravitino and Polonyi problems are avoided, and super heavy gravitinos can 
account for the correct amount of Cold Dark  Matter (CDM). The dominating channel 
of gravitino production is due to decays of Polonyi particles,  in turn, produced during inflation. 
On the other hand, we found that direct production of gravitinos 
during inflation is a subleading process that does not significantly change our estimates. 

Intriguingly, our results imply that the parameter spaces of cold DM and inflation 
can be linked to each other,  into a natural unifying picture.  This emerging DM picture 
suggests a phenomenology  in ultra high energy cosmic rays.  For example, super heavy Polonyi particles 
can decay into the SM  particles as secondaries in  top-bottom decay processes. Cosmological high energy neutrinos  from primary and secondary channels  may be detected in IceCube and ANTARES experiments. 
A numerical investigation of these channels  deserves further investigations, beyond the purposes of this paper. 

Our scenario offers a link to the realistic Supersymmetric Grand Unified Theories (SGUTs) coupled to
supergravity.  The super-Higgs effect considered in Sec.~2 is associated with the $U(1)$ gauge-invariant 
supersymmetric field theory. This  $U(1)$ can be naturally embedded into a SGUT with a non-simple gauge group.  The Starobinsky inflationary scale, defined by either $m_{\rm inf}$ or $H_{\rm inf}$ is by three or two orders of magnitude lower, respectively, than the SGUT scale of $10^{16}$ GeV.  The SGUTs with the simple gauge group $SU(5)$, $SO(10)$ or $E_6$ are well motivated beyond the Standard Model. However, the SGUTs originating from the heterotic string compactifications on Calabi-Yau spaces usually come with one or more extra $U(1)$ gauge factors as e.g., in the following gauge symmetry breaking patterns: $E_6\to SO(10)\times U(1)$,  $SO(10)\to SU(5)\times U(1)$, the "flipped" $SU(5)\times U_X(1)$ and so on (see e.g.,
Ref.~\cite{Ellis:2016spb} for more). Alternatively, SGUTs can be obtained in the low energy limit of intersecting D-branes in type IIA or IIB closed string theories.  Also in this context, extra $U(1)$ factors in the
gauge group are unavoidable.  For instance, the "flipped" $SU(5)\times U_X(1)$ from the intersecting D-brane models was studied in Refs. \cite{Antoniadis:2000ena,Blumenhagen:2001te,Ellis:2002ci,Cvetic:2002pj}.
 
We propose to identify one of the extra $U(1)$ gauge (vector) multiplets with the  inflaton vector multiplet
considered here.  This picture would be very appealing because it unifies SGUT, inflation and DM. 
Moreover, the extra $U(1)$ gauge factor in the SGUT gauge group  may  also stabilize proton and get rid of monopoles, domain walls and other topological defects \cite{Ketov:1996bm}. 

Our scenario also allows us to accommodate a positive cosmological constant, i.e. to include dark energy (see
the end of Sec.~2) too. Further physical applications of our supergravity model for inflation and DM  to SGUTs and reheating  are very sensitive  to interactions between the supergravity sector and  the SGUT fields. Demanding consistency of the full picture  including SGUT, DM and inflation may lead to further constraints.
 For instance, a matter field must be weakly coupled to inflaton ---  less then $10^{-3}$ ---
  in order to preserve the almost flat plateau of the inflaton scalar potential.  Among the other relevant issues, the Yukawa coupling of inflaton to a Right-Handed (RH) neutrino is  very much connected to  
  the leptogenesis issue.   Inflaton can also decay into RH-neutrinos,   in turn, decaying into 
SM  visible particles.  Of course, these remarks are very generic and have low predictive power, being highly model-dependent.   But they motivate us for a possible derivation of  our  supergravity model from superstrings --- see e.g., Ref.~\cite{Ellis:2016spb} for the previous attempts  along these lines. 

Another opportunity can be based on Refs.~\cite{Cvetic:2004ui,Addazi:2015yna},  by
 adapting the Pati-Salam model to become predictive in the neutrino mass sector and be
accountable for leptogenesis in supergravity, as was suggested in Ref.~\cite{Addazi:2016mtn}.
Then one can generate a highly degenerate mass spectrum of RH neutrinos, close to $10^{9}\, {\rm GeV}$, i.e. four orders smaller than the inflaton mass.  In this approach an extra $U(1)$ is necessary for consistency, while it can be related to the Higgs sector in supergravity. 

It is worth mentioning that our hidden sector includes only inflaton and Polonyi, and it may have to be extended. The scale invariance of the (single-field) Starobinsky inflation is already broken by the mixing of inflaton with Polonyi scalar, while its breaking is necessary for  a formation of mini-PBHs during inflation. The physical consequences of the inflaton-Polonyi mixing demand a more detailed investigation, beyond the scope of this paper. 

Our scenario can be reconsidered in the general framework of Split-SUSY and High-scale SUSY, by questioning its compatibility with the SM and the known Higgs mass value of 125.5 GeV in particular, e.g., along the lines of the comprehensive study in Ref.~\cite{Giudice:2011cg}.~\footnote{Our scenario is apparently incompatible with 
Low- or Intermediate-scale SUSY that imply a significantly lower gravitino mass and a substantial inflaton decay rate into
gravitino pairs, see e.g., Ref.~\cite{Terada:2014uia} for details.}
 Then the "unification help" from SUSY to GUT scenarios could be  implemented in our model.  In this case, several new decay channels are opened and new parameters enter. In particular, there is a scenario in which the Higgsino at $1-100\, {\rm TeV}$ scale is envisaged, with intriguing implications for future colliders. In such case, produced gravitinos can decay into Higgsinos that (in the form of neutralinos) could provide another candidate for Dark Matter. However, there also exist contributions from thermal production that may affect the Higgsino production.  Actually, the upper bound on the scale of Split-SUSY, according to Fig.~3 of Ref.~\cite{Giudice:2011cg}, is given by $10^8$ GeV that already excludes compatibility of Split-SUSY with our scenario that requires a higher SUSY. In the case of High-scale SUSY, the upper bound in Fig.~3 of Ref.~\cite{Giudice:2011cg} is given by 
$10^{12}$ GeV for a considerable part of the parameter space, so that this bound  is again too low for our model. However,  
it is still possible to go beyond that bound in the case of High-scale SUSY, as is shown in Fig.~5 of 
Ref.~\cite{Giudice:2011cg}, so that our SHGDM scenario is still allowed. A more detailed study of the compatibility of our scenario with the SM deserves further investigation, beyond the scope of this paper. 

\section*{Appendix: the power spectrum of Polonyi and gravitino emissions }

Below we provide some technical details about our calculation of the power spectrum of Polonyi and gravitino emissions,
based on finding a numerical solution to  Eq.~(\ref{modes}) in the framework of adiabatic  theory  \cite{Adiabatic1,Adiabatic2,Adiabatic3}. 

A change of variables from $h_{k}$  in Eq.~(\ref{XX}) to $W_{k}$ as
\begin{equation}
\label{WKB}
h_{k}=(\sqrt{2W_{k}})^{-1}{\rm exp}\left( -i\int^{\eta}W_{k}(\eta')d\eta'\right)~,
\end{equation}
and plugging this into Eq.~(\ref{XX}) yield  
\begin{equation}
\label{diff}
W_{k}^{2}=w_{k}^{2}-\left[W_{k}''/2W_{k}-\frac{3}{4} \left(W_{k}'/W_{k}\right)^{2}  \right]\, ,
\end{equation}
where $'=\partial/\partial \eta$ denotes the derivative with respect to the conformal time.  

Applying the Bogoliubov transformation to $h_{k}$ leads to the following relation among $W_{k}$ and $\beta$:
\begin{equation}
\label{beta}
|\beta_{k}(\eta_{1},\eta_{0})|^{2}=(4W_{k}^{\eta_{0}}\,W_{k}^{\eta_{1}})^{-1}\{ (\zeta_{k}'\,^{\eta_{0}}-\zeta_{k}'\,^{\eta})^{2}+(W_{k}^{\eta_{0}}-W_{k}^{\eta_{1}})^{2}\}\, , 
\end{equation}
where 
\begin{equation}
\label{deak}
\zeta_{k}'^{\,\eta}=W_{k}'\,^{\eta}/2W_{k}^{\eta}\, . 
\end{equation}

The adiabatic approximation consists of considering  the background metric to be slowly changing
in time, so that the time variation can be treated by introducing a small parameter $\epsilon$ via the substitution 
$\partial/\partial \eta\rightarrow \epsilon \partial/\partial \eta$ (this approximation can be reasonably applied during inflation),
\begin{equation}
\label{EEE}
W=W^{(0)}+\epsilon^{2}W^{(2)}+\epsilon^{4}W_{(4)}+ \ldots~,
\end{equation}
where the label $(n)=(0),(2),(4),...$ denotes the adiabatic order expansion. The numerical problem can be solved by defining the iterative map as follows: 
\begin{equation}
\label{Mapp}
\mathcal{M}[W_{k}^{n}]=\sqrt{w_{k}^{2}-\frac{1}{2}\left[W_{k}''^{(n)}/W_{k}^{(n)}-\frac{3}{2} \left(W_{k}'^{(n)}/W_{k}^{(n)}\right)^{2}  \right]}~.
\end{equation}
This map raises the adiabatic order as 
\begin{equation}
\label{Wn2}
W_{k}^{(n+2)}=\mathcal{M}[W_{k}^{(n)}]~,\quad W_{k}^{(0)}=w_{k}~.
\end{equation}

A few leading terms of the adiabatic order expansion can be straightforwardly calculated, with the following results: 
\begin{equation}
\label{WW}
W^{(0)}=w,\,\,\, W^{(2)}=3w'^{2}/8w^{3}-w''/4w^2~,\nonumber
\end{equation}
\begin{equation}
\label{WWW}
W^{(4)}=-k_{1}(w')^{4}/w^{7}+k_{2}(w')^{2}w''/w^{6}-k_{3}(w'')^{2}/w^{5}-k_{4}w'w'''/w^{5} +k_{5}w'''/w^{4}~,
\end{equation}
where 
$k_{1}=297/128$, $k_{2}=99/32$, $k_{3}=13/32$, $k_{4}=5/8$ and $k_{5}=1/16$.  The  $j$-th adiabatic order reads 
\begin{equation}
\label{adiabaticA}
h_{k}^{(j)}=(1/\sqrt{2W_{k}^{j}})\,{\rm exp}\left( -i\int^{\eta}W_{k}^{(j)}(\eta')d\eta'\right)~,
\end{equation}
where the adiabatic vacuum state of the j-th order is defined by specifying the boundary conditions at a fixed value $\eta^*$ of $\eta$ as
\begin{equation}
\label{hhh}
h_{k}(\eta^{*})=h_{k}^{(j)}(\eta^{*}),\,\,\, h_{k}'(\eta^{*})=h'^{(j)}_{k}(\eta^{*})~.
\end{equation}

A similar iterative procedure can be applied to gravitino also. The $b_{\mu}$ modes can be decomposed 
into two vector-spinor fields,
\begin{equation}
\label{twoo}
b_{k\mu}=(h_{k}^{I} u_{\alpha \mu}, h_{k}^{II} u^{\dot{\alpha}^{\dagger}}_{\mu})^{T}~,
\end{equation}
where $u$ is a vector-spinor of spin 3/2, while $h^{I,II}$ has the structure similar to that of Eq.~(\ref{WKB}) in terms of other functions $W_{\pm}$. The equations of motion for $W_{\pm}$ are very complicated,
\begin{equation}
\label{WKC}
[-W_{\pm}^{3/2}+3W'^{2}_{\pm}/(4W^{5/2}_{\pm}) -W''_{\pm}/(2 W^{3/2}_{\pm})]-i(\gamma^{\mu}\gamma^{i}+\gamma^{i}\gamma^{\mu})k_{i}(\gamma_{\nu}\gamma_{0}+\gamma_{0}\gamma_{\nu})[\sqrt{W}_{\pm}-W'_{\pm}/2W^{3/2}_{\pm}]/2 \nonumber
\end{equation}
$$-2\gamma_{\nu}(m_{3/2}+i\frac{a'}{a}\gamma^{0})i(\gamma_{\nu}\gamma_{0}+\gamma_{0}\gamma_{\nu})[\sqrt{W_{\pm}}-W'_{\pm}/2W_{\pm}^{3/2}]+(k^{2}+m_{3/2}^{2}+2i\frac{a'}{a}\gamma^{0}m_{3/2})(2W_{\pm})^{-1/2}~,$$
\begin{equation}
\label{WKC1}
[-W_{-}^{3/2}+3W'^{2}_{-}/(4W^{5/2}_{-}) -W''_{-}/(2 W^{3/2}_{-})]+i(\bar{\sigma}_{\mu}\sigma_{i}+\bar{\sigma}_{i}\sigma_{\mu})k_{i}(\bar{\sigma}^{\nu}\sigma^{0}+\bar{\sigma}^{0}\sigma^{\nu})[\sqrt{W}_{-}-W'_{-}/2W^{3/2}_{-}]/2
\end{equation}
$$-2\sigma^{\nu}(m_{3/2}+i\frac{a'}{a}\sigma_{0})i(\bar{\sigma}^{\nu}\sigma^{0}+\bar{\sigma}^{0}\sigma^{\nu})[\sqrt{W_{-}}-W'_{-}/2W_{-}^{3/2}]+(k^{2}+m_{3/2}^{2}+2i\frac{a'}{a}\sigma^{0}m_{3/2})(2W_{-})^{-1/2}~,$$
\begin{equation}
\label{WKC2}
[-W_{+}^{3/2}+3W'^{2}_{+}/(4W^{5/2}_{+}) -W''_{+}/(2 W^{3/2}_{+})]-i(\sigma^{\mu}\bar{\sigma}^{i}+\sigma^{i}\bar{\sigma}^{\mu})k_{i}(\sigma_{\nu}\bar{\sigma}_{0}+\sigma_{0}\bar{\sigma}_{\nu})[\sqrt{W}_{+}-W'_{+}/2W^{3/2}_{+}]/2 \nonumber
\end{equation}
$$-2\sigma_{\nu}(m_{3/2}+i\frac{a'}{a}\sigma^{0})i(\sigma_{\nu}\bar{\sigma}_{0}+\sigma_{0}\bar{\sigma}_{\nu})[\sqrt{W_{+}}-W'_{+}/2W_{+}^{3/2}]+(k^{2}+m_{3/2}^{2}+2i\frac{a'}{a}\gamma^{0}m_{3/2})(2W_{+})^{-1/2}~, \nonumber$$
where we have used the notation $\bar{\sigma}=(I,-\sigma_{i})$ and $\sigma=(I,\sigma_{i})$.

One arrives at the following gravitino spectrum:
\begin{equation}
\label{beta}
|b_{k\mu}(\eta_{1})b^{C \mu}_{k}(\eta_{0})|=(4W_{k\,-}^{\eta_{0}}\,W_{k\,+}^{\eta_{1}})^{-1}\{ (\zeta_{k\,-}'^{\eta_{0}}-\zeta_{k\, +}'^{\eta_{1}})^{2}+(W_{k\, -}^{\eta_{0}}-W_{k\, +}^{\eta_{1}})^{2}\}~,
\end{equation}
with a similar equation for Polonyi fields. Plugging in the adiabatic solutions for $\beta$'s and $b$'s into the above equations leads to the
final power spectra, though only numerically.  

Our numerical results can be compared with the semi-analytical results of Ref.~\cite{Chung:1998bt}, where the steepest descent method of integration was used, leading to the crude estimate
\begin{equation}
\label{betajj}
|\beta_{k}|^{2}\simeq {\rm exp}\left[ -4 M_{A}(H_{i}^{2}+R_{i}/6)^{-1/2})\right]~,
\end{equation}
where $H_{i}$ and $R_{i}$ are the Hubble parameter and the Ricci scalar during inflation. Inserting $R_{i}=6(\ddot{a}/a+H_{i}^{2})\simeq H_{i}^{2}$, $H_{i}=1.4\times 10^{14}\, {\rm GeV}$  and $M_{A}=1.5 \times 10^{13}\, {\rm GeV}$ into (\ref{betajj}) yields $|\beta_{k}|^{2}\simeq 0.67$,  in basic agreement with our numerical results. It is worth noticing that, according to Ref.~\cite{Chung:1998bt}, Eq.~(\ref{betajj}) is subject to ${\cal O}(k/M_{A})$ corrections.
\vglue.2in
  
\section*{Acknowledgements}

S.V.K. was supported by a Grant-in-Aid of the Japanese Society for Promotion of Science (JSPS) under No.~26400252, the Competitiveness Enhancement Program of Tomsk Polytechnic University in Russia, and the World Premier International Research Center Initiative (WPI Initiative), MEXT, Japan. S.V.K. would like to thank a FAPESP grant 2016/01343-7 for supporting his visit in August 2017 to ICTP-SAIFR in Sao Paulo, Brazil,  where part of this work was done. The work by M.K. was supported by Russian Science Foundation and fulfilled in the framework of MEPhI Academic Excellence Project (contract № 02.a03.21.0005, 27.08.2013). The authors are grateful to Y. Aldabergenov, A. Marciano, K. Hamaguchi, A. Kehagias, K. Kohri, A. A. Starobinsky and T. Terada for discussions and correspondence.

\bibliographystyle{utphys} 

\providecommand{\href}[2]{#2}\begingroup\raggedright
\endgroup

\end{document}